\documentstyle[12pt]{article}

\newcommand{\be}{\begin{equation}}
\newcommand{\ee}{\end{equation}}

\begin{document}
\author{A.E. Dom\'{\i}nguez\footnote{Electronic addresss: domingue@fis.uncor.edu} and D.E. 
Barraco\footnote{Electronic address: barraco@famaf.unc.edu.ar}  \\FaM.A.F., Universidad 
Nacional
de C\'ordoba \\
Ciudad Universitaria, C\'ordoba 5000, Argentina}  

\title{ Newtonian limit  of the singular $f(R)$ gravity in the 
Palatini formalism.} 
\maketitle

\begin{abstract}
Recently D. Vollick [Phys. Rev. D{\bf68}, 063510 (2003)] has shown that the inclusion of the $1/R$ 
curvature terms in the gravitational action and the use of the Palatini 
formalism offer an alternative explanation for cosmological acceleration.
In this work we show not only that this model of Vollick does not have a 
good Newtonian limit, but also that any $f(R)$ theory with a pole of 
order $n$ in $R=0$ and $( d^2f/d^2R) (R_0) \neq 0$, where $R_0$ is 
the scalar curvature of background, does not have a good Newtonian limit. 
\end{abstract}
PACS number(s): 95.35.+d, 04.25.Nx, 98.80.-k, 04.20.-q

\section{INTRODUCTION}\label{sec:Intro}

From recent studies it seems well established that our 
universe expansion is currently in an accelerating phase. The evidence of
cosmic acceleration has arisen not only from the high redshift surveys of 
type Ia supernovae \cite{Riess,Prelmutter98}, but also from the 
anisotropy power spectrum of the cosmic microwave background 
\cite{Bennet,Netterfield}. One of the most accepted explanations is that 
the Universe has been dominated by some form of dark energy for a long time. 
However, none of the existing dark energy models are completely 
satisfactory. 

It is possible to find other explanations for cosmic expansion using 
field equations other than Eisntein's equations. Recently, some authors 
have proposed to add a $R^{-1}$ term in the Einstein-Hilbert action to 
modify general relativity 
\cite{Carroll0306438,Capozziello1,Capozziello0303041}. They obtained the 
field equations using second-order formalism, varying only the metric 
field, and thus obtained the so-called fourth-order field equations. 
Although the models were obtained using corrections of the Einstein-Hilbert 
Lagrangian of type $R^n$, where $n$ can take a positive or negative 
value to explain both the inflation at an early time and the expansion at 
the present time \cite{Carroll0306438,Nojiri0307288}, they still suffer from 
violent instabilities \cite{Dolgov0307285}. The Newtonian limit of these fourth- 
order theories has been studied by Dick \cite{Dick0307052}.

On the other hand, we can consider those theories that are obtained from 
a Lagrangian density ${\cal L}_{T}(R) = f (R) \sqrt{-g} + {\cal L}_{M}$, 
which depends on the scalar curvature and a matter Lagrangian that does 
not depend on the connection, and then we can apply Palatini's method to 
obtain the field equations \cite{hamity,barracothesis}. In Refs. 
\cite{hamity,barracothesis}, we showed the universality of the Einstein 
equation using a cosmological constant. More recently, Ferraris, Francaviglia and 
Volovich  published the same result \cite{Francaviglia,Francaviglia1}. 
For these theories we have studied the conserved quantities \cite{barraco}, the
spherically symmetric solutions \cite{barraco1999}, the Newtonian limit 
\cite{hamity,Barr0}, and the cosmology described by Friedmann-Robertson-Walker (FRW)
metric \cite{Barra2002}.
 
In our previous paper \cite{Barr0}, it was also shown that it is very 
difficult to test these models in the (post-)Newtonian approximation. The 
reason for this is that the departures from Newtonian behavior are both 
very small and are masked by other effects, because these departures from the 
Newtonian behavior have to be measured when the body is moving ``through" 
a matter-filled region. In the main applications of these theories, for 
example, in the Newtonian limit or in the calculation of the observed 
cosmological parameter, we have considered the analyticity  of $f(R)$ at 
$R=0$. 

Moreover, it is well known \cite{Barr0,Francaviglia,barraco1999} that in a 
vacuum, or in the case of $T=const$, the solutions of these theories 
are the same as general relativity with a cosmological constant, even when 
$ f(R)$ is not analytical at $R=0$. On the other hand, solutions 
corresponding to different cosmological constants are allowed by some of 
these theories. Therefore, the homogeneous and isotropic vacuum solution 
for these theories is the de Sitter space-time with different 
cosmological constants, except when one of the  allowed cosmological 
constants is $\Lambda=0$, which corresponds to flat space-time.

Recently, Vollick \cite{vollick} used the corrected Lagrangian of the 
works of Carroll {\it {\it et al}.}, and Capozziello {\it {\it et al}.} 
\cite{Carroll0306438,Capozziello0303041}, $ f(R) = R - {\alpha}^2/ R$, and 
the Palatini variational principle --which is a particular case of the 
above theories but is not analytical at $R=0$-- to prove that the 
solutions of the field equations approach de Sitter universe at a late 
time. This result was obtained using the above well known property of the 
vacuum solutions. Thus, the inclusion of  $1/R$ curvature terms in the 
gravitational Lagrangean provides us with an alternative explanation for 
the cosmological acceleration. Moreover, Meng and Wang 
\cite{Meng0308031,Meng0307354} have studied the modified Friedmann 
equation with the Palatini variational principle and its first-, second-, and 
third-order approximated equations.

In agreement with this last idea about the maximal symmetric vacuum 
solution, we have studied the Newtonian limit for $f(R)$ theories as a 
perturbation of the de Sitter background. The difference from previous 
works \cite{hamity,Barr0} is that there the background was flat 
space-time, which is only possible in the particular case  $f(0)=0$. 

In this work we show that the Vollick model \cite{vollick} does 
not give the Newtonian limit but also that any $f(R)$ with a pole of 
order $n$ in $R=0$ and $( d^2f/d^2R) (R_0) \neq 0$, where $R_0$ is 
the scalar curvature of background, cannot have a Newtonian Limit. This 
result is equivalent to the condition obtained by Dick 
\cite{Dick0307052} for the fourth-order theories. 
Nevertheless, the theories with singular $f(R)$ and $ (d^2f/d^2R) 
(R_0) = 0$ must be studied in the future since they have a correct 
Newtonian limit and explain the observed cosmological acceleration.
In our work we follow the conventions and notation of Synge \cite{Synge}
\section{REVIEW OF THE FIELD EQUATIONS} \label{sec:Struc}

We review in this section the structure of the theory as presented
elsewhere in previous works \cite{hamity, Barr0}.

Let $\cal{M}$ be a manifold with metric $g_{ab}$ and a torsion-free 
derivative operator $\nabla_{a}$, both considered as independent 
variables. Consider a  Lagrangian density $ {\cal L} =  f(R)\sqrt{-g} + 
{\cal L}_{M}$, where the matter Lagrangian ${\cal L}_M$ does not depend 
on 
the connection. 

Suppose  we have a smooth one-parameter $\lambda$ family of field 
configurations starting from given fields $g^{ab}$, $\nabla_{a}$, and 
$\psi$ (the matter fields), with appropriate boundary conditions.  
Let $\delta g^{ab}$, $\delta \Gamma^{c}_{ab}$, $\delta \psi$ be the 
corresponding variations of those fields, i.e., $\delta g^{ab} = 
(dg_{\lambda}^{ab}/d\lambda)\mid_{\lambda=0}$, etc. 
 Then, if we vary with respect to the metric, the field equations are
\be
f'(R) R_{ab} - \frac{1}{2} f(R) g_{ab} = \alpha_M T_{ab}. \label{eq:1}
\ee
where $f'(R) = (df/dR)$, $(\delta S_{M}/\delta g^{ab}) \equiv - T_{ab} 
\sqrt{-g}$ and $\alpha_m = -8\pi$.
The variation with respect to the connection, recalling that this is 
fixed 
at the boundary, gives
\be
\nabla_c[\sqrt{-g} g^{ab} f'(R)]=0. \label{eq:2}
\ee
Now, we choose Lagrangian $f(R)$ with $f'(R)$ derivable. Then, the last 
equation becomes  
\be
\nabla_{c} g_{ab} = b_{c} g_{ab}, \label{eq:3}                              
\ee
where
\be
b_{c} = - [\ln  f'(R)]_{,c}.  \label{eq:4}
\ee
Thus, we have a Weyl conformal geometry with a Weyl field given by 
Eq. (\ref{eq:4}).

From Eq. (\ref{eq:1}) we obtain 
\be
f'(R) R - 2 f(R) = \alpha_M T,   \label{eq:5}
\ee
which defines $R(T)$, and we suppose that the function $f(R)$ is such that 
$R(T)$ is derivable with respect to the variable $T$. Therefore, $b_{c}$ 
is determined by $T$ and its derivatives except in the case $ f(R) = 
\omega R^2$, for which $R f'- 2 f \equiv 0$, so we must 
consistently have $ T \equiv 0$. 
It is important to note that $b_{c}$ has a solution only if $T$ is
differentiable in  $\cal{M}$; this condition on $T$, for the existence of a solution, is 
not necessary in other theories such as general relativity (GR) or fourth-order theories.

The connection solution to Eq. (\ref{eq:3}) is 
\be
\Gamma^{a}_{bc} = C^{a}_{bc} - \frac{1}{2} ( \delta^{a}_{b} b_{c} + 
\delta^{a}_{c} b_{b} - g_{bc} b^{a}),  \label{eq:6}
\ee
where $C^{a}\,_{bc}$ are the Christoffel symbols (metric connection). 
Then, we have to solve only Eq. (\ref{eq:1}).

The Riemann tensor can be defined in the usual form, and then, the Ricci 
tensor and scalar curvature are
\be
R_{ab} = R^{m}_{ab} -\frac{3}{2} D_{a}b_{b} + \frac{1}{2} D_{b} b_{a} - 
\frac{1}{2} g_{ab} D\cdot b - \frac{1}{2} b_a b_b + \frac{1}{2} g_{ab} 
b^2 
\label{eq:7} 
\ee
\be
R = R^{m} - 3 D\cdot b + \frac{3}{2} b^2, \label{eq:8} 
\ee
where $R^{m}_{ab}$, $R^m$, and $D_c$ are the Ricci tensor, scalar 
curvature, and covariant derivative defined from the metric connection, 
respectively.

Because the matter action must be invariant under diffeomorphisms and the 
matter fields satisfy the matter field equations, $ T_{ab}$ is 
conserved \cite{Barr0} 
\be
D^{a}T_{ab} = 0.  \label{eq:10}
\ee
Therefore,  a test particle will follow the geodesics 
of the metric connection.
Using Eqs.(\ref{eq:4}) and (\ref{eq:5}) we have
\be
b_{c} = - \frac{f''\alpha_M \nabla_{c} T}{f'(R f''-f')}. \label{eq:14}
\ee
 Except for the case of GR, where $f'' \equiv 0$, the Weyl field is 
nonzero 
wherever the trace of the energy-momentum tensor varies with respect to 
the coordinates. If $T$ is constant, then $R$ is also constant, $b_{c} 
=0$ 
and (\ref{eq:5}) takes the form 
\be
G_{ab} -\frac{1}{2} \Lambda g_{ab} = K T_{ab}, \label{eq:15}
\ee
where $\Lambda$ and $K$ are two functions of $R$. 
All those cases with constant trace of the energy-momentum tensor are 
equivalent to GR for a given cosmological constant. This is the so-called 
universality of the Einstein equations for matter for which  
$T$ is constant \cite{hamity,barracothesis}. In the case $T=0$, the scalar $R$
is any of the roots, $R_{i},$ of the equation $f'(R) R -2 f(R)=0$. For each 
root the solutions of the field equations are the solutions of GR with cosmological
 constant $\Lambda= -R_{i}/4$. Therefore, the maximal symmetric vacuum solution of 
these theories is the de Sitter space-time.

\section{THE NEWTONIAN LIMIT}

 The above theories, which explain the cosmological acceleration, must be 
checked in the\\ Newtonian limit, i.e., in the slow field and slow velocity 
approximation. In this approximation, these theories must be in agreement 
with the present data. 

The Newtonian limit, in our case, must be taken as a perturbation of the 
homogeneous isotropic vacuum background. 
As we have seen above, this background space-time is the de Sitter space-time,
 except when $ \Lambda=0$, in which case we must perturbate around 
the flat space-time. The case $\Lambda =0$ has just been studied 
\cite{hamity,Barr0} and the general solution of the problem with $\Lambda 
\neq 0$ must agree with the previous result when $ \Lambda =0 $. 

The background metric and the background Ricci tensor are
\be
\stackrel{0}{g}_{ab} = - dt^2 + e^{2 t\sqrt{\Lambda/3} } ( 
dx^2+dy^2+dz^2), \label{eq:16}
\ee
\be
\stackrel{0}{R}_{ab} = -\Lambda \stackrel{0}{g}_{ab}. \label{eq:17}
\ee 
We assume that the metric can be written in the form
\be
g_{ab}= \stackrel{0}{g}_{ab} + h_{ab} \label{eq:18}
\ee
where $h_{ab}$ is the perturbation of the metric.
The first order of the field equations (\ref{eq:1}) are 
\be
\stackrel{1}{f'}(R)\stackrel{0}{R}_{ab} + 
\stackrel{0}{f'}(R)\stackrel{1}{R}_{ab} - \frac{1}{2} \stackrel{1}{f}(R) 
\stackrel{0}{g}_{ab} - \frac{1}{2}\stackrel{0}{f}(R) h_{ab} = \alpha_M 
\stackrel{1}{T}_{ab} \label{eq:19}
\ee 
where the $0$ superscript means zero-order quantities, i.e., background 
quantities, and the superscript $1$ means the first-order quantities. We have to 
calculate each term of the above equation.

According to (\ref{eq:6}) the  connection $\Gamma^c_{ab}$ can be split 
into  two parts,
\be
\Gamma^c_{ab} = \stackrel{0\;}{ \Gamma^c}_{ab} + 
\stackrel{1\;}{\Gamma^c}_{ab}, \label{eq:20}
\ee
where $\stackrel{0\;}{ \Gamma^c}_{ab} $ is the connection corresponding 
to the background metric  and $\stackrel{1\;}{\Gamma^c}_{ab}$ has two 
terms,  one  depending on the metric perturbation $h_{ab}$ and the other 
depending on $b_a$ which has an order higher than zero.                     

In order to be able to write the linear  field equations we must 
calculate the linear Ricci tensor by using (\ref{eq:20}). We obtain
\be
R_{ab} = \stackrel{0}{R}_{ab} + \stackrel{1}{R}_{ab} \label{eq:21}
\ee
where the$\stackrel{1}{R}_{ab}$ part is
\be
\stackrel{1}{R}_{ab} = - \stackrel{0}{\nabla}_c 
\stackrel{1\;}{\Gamma^c}_{ab} + \stackrel{0}{\nabla}_{(a} 
\stackrel{1\;}{\Gamma^c}_{b)c}. \label{eq:22}
\ee
By $\stackrel{0}{\nabla}$ we mean the covariant derivative associated with the 
background metric.

As we have just said, the first order of connection can be written as 
follows:
\be
\stackrel{1\;}{\Gamma^c}_{ab} = \stackrel{1\;}{C^c}_{ab} + 
\stackrel{1\;}{A^c}_{ab}, \label{eq:23}
\ee 
where the first term is
\be
\stackrel{1\;}{C^c}_{ab} = 
\stackrel{0\;\;\;}{g^{cd}}\stackrel{0}{\nabla}_{(a}h_{b)d}-\frac{1}{2} 
\stackrel{0\;\;\;}{g^{cd}}\stackrel{0}{\nabla}_d h_{ba} \label{eq:24}
\ee
and the second term is
\be
\stackrel{1\;}{A^c}_{ab} = - \frac{1}{2} ( \delta ^c_a \stackrel{1}{b}_b 
+ \delta ^c _b \stackrel{1}{b}_a-\stackrel{0}{g}_{ab}\stackrel{1\;}{b^c}). 
\label{eq:25}
\ee

By substituting (\ref{eq:23}) in (\ref{eq:22}) we obtain
\be
\stackrel{1}{R}_{ab} = 
\stackrel{0}{\nabla}_{(a}\stackrel{1\;}{C^c}_{b)c}-\stackrel{0}{\nabla}_c 
\stackrel{1\;}{C^c}_{ab} + 
\stackrel{0}{\nabla}_{(a}\stackrel{1\;}{A^c}_{b)c} - 
\stackrel{0}{\nabla}_c \stackrel{1\;}{A^c}_{ab}. \label{eq:26}
\ee

In turn, straightforward calculations lead us to

\be
\stackrel{0}{\nabla}_{(b} \stackrel{1\;}{C^c}_{a)c} = \frac{1}{2} 
\stackrel{0}{\nabla}_a \stackrel{0}{\nabla}_b\;h, \label{eq:26'}
\ee
\be
\stackrel{0}{\nabla}_c \;\stackrel{1\;}{C^c}_{ab} = 
\stackrel{0}{\nabla}_{(a} \stackrel{0\;}{\nabla^c} h_{b)c} + 
\stackrel{0\;\;\;\;\;\;\;}{R_{(ba)}^{e\;\;\;\;c}} h_{ec} - 
\stackrel{0}{R}_{e(a}h_{b)e} - \frac{1}{2} \stackrel{0}{\Box} h_{ab}. 
\label{eq:27}
\ee

Hence, by replacing Eqs. (\ref{eq:26'}) and (\ref{eq:27}) in 
(\ref{eq:26}) we obtain
\begin{eqnarray}
\stackrel{1}{R}_{ab} = \frac{1}{2} \stackrel{0}{\nabla}_a 
\stackrel{0}{\nabla}_b h - 
\stackrel{0}{\nabla}_{(a}\stackrel{0}{\nabla^c} h_{b)c} - 
\stackrel{0\;\;\;\;\;}{R_{(ba)}^{e\;\;\;\;c}} h_{ec} + 
\stackrel{0\;}{R^e}_{(a} h_{b)e} \nonumber \\ 
+ \frac{1}{2} \stackrel{0}{\Box} h_{ab}-2 
\stackrel{0}{\nabla}_{(a}\stackrel{1}{b}_{b)} 
+\frac{1}{2}(\stackrel{0}{\nabla}_a \stackrel{1}{b}_b 
+\stackrel{0}{\nabla}_b \stackrel{1}{b}_a - \stackrel{0}{g}_{ab} 
\stackrel{0}{\nabla}_c \stackrel{1\;}{b^c}). \label{eq:28}
\end{eqnarray}

As is well known, there is a gauge freedom in any geometrical theory 
of gravitation corresponding to the group of diffeomorphisms of 
space-time. In practice, these diffeomorphisms may be viewed as coordinate 
freedom which may be used to impose coordinate conditions. For instance, 
we may choose coordinates $x^a$ so that, in the linear approximation, 
the perturbation $h_{ab}$ and  the vector field $b_a$ satisfy the 
gauge condition
\be
\stackrel{0\;}{\nabla^c} h_{bc} - \frac{1}{2}\stackrel{0}{\nabla}_b h + 
\stackrel{1}{b}_b = 0. \label{eq:30}
\ee
In this gauge the linearized Ricci tensor simplifies to become
\be
\stackrel{1}{R}_{ab} = \frac{1}{2} \stackrel{0}{\Box} h_{ab} - 
\stackrel{0\;\;\;\;\;\;}{R_{(ba)}^{e\;\;\;\;c}} h_{ec} + 
\stackrel{0\;}{R^e}_{(a}h_{b)e} - \frac{1}{2} \stackrel{0}{g}_{ab} 
\stackrel{0}{\nabla}_c \stackrel{1\;}{b^c}. \label{eq:31}
\ee

Finally, to rewrite the linear field equations (\ref{eq:19}) we have to 
take into account the factors $\stackrel{1}{f'}(R)$ 
and $\stackrel{0}{f}(R)$ which depend on $T$. From (\ref{eq:21}) we can 
split the curvature scalar into a zero order part and a first order part:
\be
R=R_0 + \stackrel{1}{R}, \label{eq:32}
\ee
where $ R_0 = -4 \Lambda$.
Therefore, using Eq. (\ref{eq:5}) for first terms we can obtain 
$\stackrel{1}{R}$ as a function of $\stackrel{1}{T}$:
\be
\stackrel{1}{R} = \frac{\alpha_M \stackrel{1}{T}}{f''(R_0) R_0 - f'(R_0)}. 
\label{eq:33}
\ee
From this result we can easily obtain $\stackrel{0}{f}(R)$ and 
$\stackrel{1}{f'}(R)$.

Now we are ready to rewrite the first order field equations, in the gauge 
(\ref{eq:30}), using Eqs. (\ref{eq:14}), (\ref{eq:19}), 
(\ref{eq:31}), (\ref{eq:33}), and (\ref{eq:17}):
\begin{eqnarray}
\frac{1}{2}\stackrel{0}{\Box} h_{ab} - \frac{4 \Lambda}{3}h_{ab} 
+\frac{\Lambda}{3} h \stackrel{0}{g}_{ab}- \frac{1}{2} \frac{\alpha_M 
\stackrel{0}{f''}}{\stackrel{0}{f'}(\stackrel{0}{f''} 4\Lambda + 
\stackrel{0}{f'})} \stackrel{0}{g}_{ab} \stackrel{0}{\Box} 
\stackrel{1}{T} \nonumber\\-\frac{1}{2} 
\frac{\stackrel{0}{f}}{\stackrel{0}{f'} }h_{ab}=  
\frac{\alpha_M}{\stackrel{0}{f'}} \stackrel{1}{T}_{ab} - \left[ 
\frac{\alpha_M}{2(4\stackrel{0}{f''}\Lambda + \stackrel{0}{f'})}\left( 
1+\frac{\stackrel{0}{2 
f''}\Lambda}{\stackrel{0}{f'}}\right)\right]\stackrel{1}{T} 
\stackrel{0}{g}_{ab}. \label{eq:34}
\end{eqnarray}
It is important to note that when $\Lambda = 0 $ we recover the result 
obtained in Refs. \cite{hamity,Barr0}.

In the Newtonian limit, the equation of motion of a test particle is 
given by (\ref{eq:10}) with $ u^{\alpha} \simeq \delta^{\alpha}_0$, and 
the proper time of the particle may be approximated by the coordinate 
time, $t$. Since the sources are "slowly varying", we expect the space-
time geometry to change slowly as well, so that the time derivative can 
be assumed to be negligible. 
Thus, we find
\be
\Gamma _{00}^{\mu} = -\frac{1}{2} e^{2 t \sqrt{\Lambda/3}} h_{00,\mu}, 
\;\;\;\;\;\mu=1,2,3 .\label{eq:35} 
\ee                                                                                                                                                                                                                                                                                                                                                                                                                                                                                                               

As a consequence, we have to study the Newtonian limit of $h_{00}$ using Eq. (\ref{eq:34}).
Taking into account the expressions for  $\stackrel{0}{\Box} h_{00}$ and 
$\stackrel{0}{\Box} \stackrel{1}{T}$, and the gauge condition 
(\ref{eq:30}) in order to substitute $ h_{12,2}$, we obtain the field 
equation for $h_{00}$ 
\be 
e^{-2t \sqrt{\Lambda /3}} \;\;\nabla^2 \left( \frac{h_{00}}{2}- A \rho \right) - B 
\;\left[ \frac{h_{00}}{2} - A \rho \right] = C \rho, 
\ee \label{eq:36}
where
\be
A= \frac{ \alpha_M  \stackrel{0}{f''} }{2 \stackrel{0}{f'} ( 4 
\stackrel{0}{f''} \Lambda + \stackrel{0}{f'})},
\label{eq:37}
\ee
\be 
C = \frac{ ( 8 \Lambda \stackrel{0}{f''} \stackrel{0}{f'} + 
(\stackrel{0}{f'})^2 )}{ (\stackrel{0}{f'})^2 ( 4 \Lambda 
\stackrel{0}{f''} + \stackrel{0}{f'} )}. \label{eq:38}
\ee
Finally, the candidate to be the Newtonian potential is
\begin{eqnarray}
\frac{h_{00}}{2}(\vec{x})= e^{2 t \sqrt{\Lambda/3}} C \int \frac 
{\rho(\vec{x}') \exp{ -( \mid \vec{x}-\vec{x}'\mid e^{t\sqrt{\Lambda 
/3}}\sqrt{2 \Lambda}\,\, )}}{\mid \vec{x}-\vec{x}'\mid} d^3x' + A 
\rho(\vec{x}).\label{eq:39}
\end{eqnarray}  

According to the observed cosmological acceleration, we have to choose 
$\Lambda \simeq 10^{-53}\; m^{-2}$. Therefore, the last result is 
consistent with our previous assumption about the time derivative. 
Moreover, the first term in (\ref{eq:39}) behaves as the Newtonian 
potential for any current experiment or observation. 

In order to ignore the second term, we have to assume  $\mid A/C \mid \ll 1 $. 
After a little algebra, this inequality can be rewritten as 
\be
\mid f''(R_0)/f'(R_0) \mid  \ll  \frac{1}{1+2R_ 0} .\label{eq:40}
\ee
When $ f(R)$ is analytical in $R=0$ we can fulfill this condition without 
problems since $R_0 \simeq 0$; but we are interested in the singular 
case, i.e. ,when $f(R)$ is singular in $R=0$. 

If $f(R)$ has a pole of order $n$ in $R=0$, the leading order singularity 
of $f''(R)$ and $f'(R)$ are $R^{-n-2}$ and $R^{-n-1}$, respectively. 
Therefore, the inequality (\ref{eq:40}) cannot be fulfilled except for 
the particular case \be
f''(R_0) =0.\label{eq:41}
\ee

The inconsistency of the Newtonian limit cannot be attributed to the 
gauge selection since the former is already present in the perturbation 
equation (\ref{eq:19}), which was derived previous to choosing the gauge. In reality, 
the problem is originated by $b_a$ whose first order of perturbation 
around $R_0$ has the factor $f''(R_0)/f'(R_0)$. In the singular case this 
factor becomes greater as $R_0$ gets smaller.

It is interesting to notice that the condition $f''(R_0)=0$ is the same 
as Dick's condition \cite{Dick0307052} for the case of the singular fourth 
order theories considered by him. This similarity is not surprising on 
account of the fact, which we will prove elsewhere, that if $ f(R)= 
R+\alpha g(R)$, then for the first-order of $\alpha$ the field equations of 
the fourth order theories and the field equations of the Palatini 
theories are the same.   

It is not difficult to show that if the Lagrangian density $ R - 
\alpha/3R$ does not satisfies the condition (\ref{eq:41}), then 
the perturbation around the maximally symmetric vacuum solution does not 
give a Newtonian limit. Nevertheless, as Dick has shown, there exist 
Lagrangian densities that satisfy this condition, for example, $ f(R)= R + 
R^2/9 \mu^2-3\mu^4/R$.    

\section{CONCLUSIONS}
We have studied the first order of perturbation from the maximally 
symmetric vacuum solutions, de Sitter space-time, in a family of theories 
which are obtained by using the Palatini formalism on a general 
Lagrangian $f(R)$.

We have proved that the  first order of perturbation does not give a good 
Newtonian limit when $f(R)$  has a pole of order "$n$" in $R=0$, and 
$f''(R_0)\neq 0$. However, except for a small correction proportional to 
$\rho$, the perturbation in the analytical case has a good Newtonian 
limit. In this analytical case we have also recovered the result of Refs. 
\cite{hamity,Barr0} when $\Lambda=0$.

While we were working on this paper Meng and Wang \cite{Meng0311019} 
claimed to have shown that the Newtonian limit is  obtained not only for 
analytical Lagrangian densities but also for the singular case. 
However, we disagree with their results. We suppose that the main problem is that 
they were not consistent with the first-order of perturbation, since they 
introduced terms $ \nabla_b\nabla_a f'(R)$ instead of using the first 
order forms of these terms, namely, $ f''(R_0)/f'(R_0)[R_0 f''(R_0)-
f'(R_0)] \nabla_b\nabla_a T$. 
Another point of disagreement, but not essential to show that the 
Newtonian limit has problems, is about the gauge. They chose the 
traceless gauge but this gauge does not exist when there are sources.     

Finally, the Lagrangian of Caroll {\it {\it et al}}., and Capozziello {\it et 
al}. \cite{Carroll0306438,Capozziello0303041} does not satisfy the 
condition $f''(R_0)=0$, and thus we have proved that the theory of 
Vollick \cite{vollick} is not in accordance with the experimental 
data. Nevertheless, it is worthwhile to study the $f(R)$ theories  in the  
Palatini formulation with an $f(R)$ singular when they fulfill condition 
(\ref{eq:41}), because they can give a good explanation of the observed 
cosmological acceleration.

\vspace{.6cm} 
{\bf  ACKNOWLEDGMENTS} 
\vspace{.4cm}

The authors would like to thank Dr. Victor Hugo Hamity for useful 
discussions and Dr. Paul Hobson for reading the manuscript. The authors also 
are very grateful to CONICET and SeCYT, Argentina, for financial support.

\end{document}